\documentclass[final,3p,times,authoryear]{elsarticle}

\usepackage{amssymb}

\usepackage{caption}
\usepackage{graphicx}
\usepackage{color, colortbl}
\definecolor{darkGray}{gray}{0.9}
\definecolor{lightGray}{gray}{0.6}
\usepackage{amsmath}

\journal{Arxiv}

\begin{document}

\begin{frontmatter}



\title{Shifts in Brain Dynamics and Drivers of Consciousness State Transitions}


\author[inst1]{Joseph Bodenheimer}
\author[inst2]{Paul Bogdan}
\author[inst3]{S\'ergio Pequito}
\author[inst1,inst4]{Arian Ashourvan}

\affiliation[inst1]{
            addressline={Department of Psychology, The University of Kansas, 1415 Jayhawk Blvd.}, 
            city={Lawrence},
            postcode={66045}, 
            state={KS},
            country={USA}}

\affiliation[inst2]{
            addressline={Ming Hsieh Department of Electrical and Computer Engineering, University of Southern California, 3740 McClintock Ave}, 
            city={Los Angeles},
            postcode={90089}, 
            state={CA},
            country={USA}}

            \affiliation[inst3]{
            addressline={Institute for Systems and Robotics, Instituto Superior Técnico, Universidade de Lisboa}, 
            city={Lisbon},
            postcode={66045}, 
            country={Portugal}}

\affiliation[inst4]{Correspondence to: Arian Ashourvan, 1415 Jayhawk Blvd, Lawrence, KS 66045,  Email:ashourvan@ku.edu}

\begin{abstract}
Understanding the neural mechanisms underlying the transitions between different states of consciousness is a fundamental challenge in neuroscience. Thus, we investigate the underlying drivers of changes during the \mbox{resting-state} dynamics of the human brain, as captured by functional magnetic resonance imaging (fMRI) across varying levels of consciousness (awake, light sedation, deep sedation, and recovery). We deploy a model-based approach relying on linear time-invariant (LTI) dynamical systems under unknown inputs (UI). Our findings reveal distinct changes in the spectral profile of brain dynamics -- particularly regarding the stability and frequency of the system's oscillatory modes during transitions between consciousness states. These models further enable us to identify external drivers influencing large-scale brain activity during naturalistic auditory stimulation. Our findings suggest that these identified inputs delineate how stimulus-induced co-activity propagation differs across consciousness states. Notably, our approach showcases the effectiveness of LTI models under UI in capturing large-scale brain dynamic changes and drivers in complex paradigms, such as naturalistic stimulation, which are not conducive to conventional general linear model analysis. Importantly, our findings shed light on how brain-wide dynamics and drivers evolve as the brain transitions towards conscious states, holding promise for developing more accurate biomarkers of consciousness recovery in disorders of consciousness.

\end{abstract}




\end{frontmatter}


\section{Introduction}
\label{sec:Introduction}

Understanding the intricate neural mechanisms governing transitions between various states of consciousness is a key challenge within neuroscience. The complexity of the brain, with its myriad interactions and dynamic states, highlights the difficulty in unraveling these processes. With the advent of modern neuroimaging techniques, particularly functional magnetic resonance imaging (fMRI), unprecedented access to whole-brain activity and its transitions between different consciousness states has emerged \citep{boly2007thoughts,coleman2009towards,lloyd2002functional,soddu2009reaching}. However, the neural underpinnings of large-scale brain dynamics and the interplay between cortical and subcortical regions across different consciousness states remain elusive.

Early investigations employing univariate analyses of functional and metabolic brain activity have revealed extensive changes in brain function across different consciousness states \citep{nakayama2006relationship}. Pioneering neuroimaging studies have demonstrated that anesthetics such as propofol produce dose-dependent, bilateral reductions in activity within the thalamus, midbrain reticular formation, cuneus-precuneus, posterior cingulate cortex, prefrontal cortices, and parietal associative cortices \citep{fiset1999brain, kaisti2003effects}. However, this univariate approach provides limited insights due to the brain's complex network interactions underlying consciousness \citep{bagshaw2013functional,crone2018linking}

Early evidence for network-level dysfunction during anesthesia emerged from two key observations. First, task-based studies revealed impaired processing in various domains, including visual \citep{heinke2001subanesthetic}, auditory \citep{heinke2004sequential, kerssens2005attenuated, plourde2006cortical}, verbal \citep{fu2005effects}, emotional \citep{paulus2005dose}, and memory \citep{sperling2002functional, honey2005ketamine}. Second, higher-order association areas, responsible for complex processing, were found to be more sensitive to the effects of anesthesia compared to lower-order regions involved in basic processing \citep{dueck2005propofol, heinke2005effects, ramani2007sevoflurane}.


Network neuroscience effectively investigates brain dynamics by examining changes in functional and structural connectivity, with the resting-state paradigm providing key insights into baseline functional activity across various consciousness states \citep{Sporns2011, luppi2021brain, crone2020systematic, demertzi2019human}. Patients in a vegetative state and minimally conscious state show decreased functional connectivity (FC) in regions related to default mode network (DMN) as well as executive control network (ECN) and auditory network when compared to healthy controls  \citep{demertzi2014multiple}. Patients showed a decrease in FC distributed in the parietal cingulate cortex, precuneus, lateral parietal cortex, and medial prefrontal cortex  \citep{wu2015intrinsic}. Studies have shown that restoration of thalamo-frontal connectivity can also serve as predictive markers for transitions toward conscious states ~\citep{crone2018restoration}. Studies using anesthesia-induced transitions also show specific brain circuits like those involving the thalamus and large-scale networks like the DMN are crucial for wakefulness. Anesthetics progressively disrupt connectivity within these networks (i.e., DMN, ECN) at higher doses, while lower-order sensory networks remain somewhat functionally connected \citep{deshpande2010altered, stamatakis2010changes, greicius2008persistent, boveroux2010breakdown}. This suggests that while basic sensory processing might persist, integrating information across brain regions is impaired under anesthesia, potentially due to disrupted subcortical thalamic regulation \citep{mhuircheartaigh2010cortical}.

While the resting-state paradigm offers insights into system-wide changes, understanding the brain's different states requires examining responses to external stimuli. Studies using transcranial magnetic stimulation show that perturbation spread varies with the conscious state, highlighting the role of thalamocortical circuitry \citep{sarasso2014quantifying}. More complex stimuli can further reveal network engagement; for example, auditory processing areas show varying activation patterns in response to musical stimuli under different levels of propofol-induced sedation \citep{dueck2005propofol}, and FC is disrupted during auditory word listening under propofol-induced sedation \citep{liu2012propofol}.



Machine learning (ML) has become a powerful tool for studying consciousness, with techniques like artificial neural networks revealing activation patterns in networks associated with wakefulness and arousal \citep{khosla2019machine,perl2020generative}. However, while these approaches excel at prediction, they often lack mechanistic explanations for how brain networks transition between states \citep{jin2023guidelines}. To bridge this gap, computational modeling approaches aim to reveal the mathematical underpinnings of neuronal activity \citep{luppi2023computational}. This allows researchers to explore how FC dynamically changes across consciousness states. Generative models describe the underlying neural activity, while dynamical models, including biophysical and phenomenological approaches, examine how changes in these systems lead to transitions in consciousness \citep{luppi2022dynamical,luppi2023computational,luppi2022whole,kandeepan2020modeling}. By combining these techniques, we can gain a deeper understanding of the intricate network dynamics that govern consciousness.

In this work, we capitalize on a publicly available dataset from \cite{Kandeepands003171,kandeepan2020modeling}, which measures resting-state dynamics in response to naturalistic auditory stimulation across different consciousness states -- wakefulness, light sedation, deep sedation, and recovery. We employ our recently developed computational framework to identify the large-scale oscillatory modes of the brain and the unknown external drivers influencing these dynamics \citep{ashourvan2022external}.  Our results demonstrate the stabilization of several oscillatory modes overlapping transmodal cortices during resting-state scans. The examination of auditory stimulation scans also reveals that these unknown inputs uncover task-specific, spatiotemporally overlapping patterns of consciousness-dependent co-activation and deactivation, which drive brain-wide dynamics. Our findings underscore the utility of this approach in characterizing brain dynamics and their responses to stimuli, providing novel insights into consciousness dynamics and potential applications in forecasting consciousness recovery, particularly in disorders of consciousness patients.


\section{Results}
\label{sec:Results}

This section explores the intricate relationship between brain activity and consciousness levels, examining how shifts from wakefulness to deep sedation manifest in the brain's dynamic oscillations and responses to external stimuli. In Section~\ref{sec:ResultsSec1}, a linear time-invariant model under unknown inputs is employed to dissect the spectral fingerprints of brain activity across various consciousness states. Building upon this, Section~\ref{sec:ResultsSec2} investigates how the brain's response to auditory stimuli changes across these states. The identified patterns of brain activity serve as a basis for classifying consciousness levels, demonstrating the potential for novel diagnostic and monitoring tools in this field.

 \subsection{LTI Systems' Spectral Fingerprints of Consciousness}
\label{sec:ResultsSec1}

\begin{figure*}
\centering
\includegraphics[width=1\linewidth ]{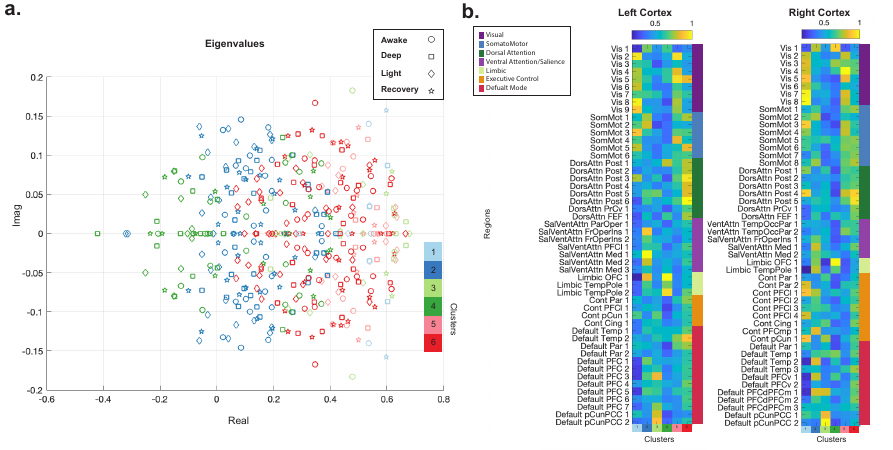}
\caption{\textbf{Spectral profiles of brain's oscillatory modes at different levels of consciousness.} \textbf{a.} Distribution of eigenvalues from \mbox{resting-state} scans across varying states of consciousness. The identified eigenvector clusters ($k=6$) are represented with distinct colors, and different consciousness states (awake, light sedation, deep sedation, and recovery) are denoted by different markers. \textbf{b.} Identified centroid of eigenvector clusters (color-coded column-wise) associated with the eigenvalue from panel \textbf{a}.}
\label{Figure_1}
\end{figure*}

To capture and describe the large-scale brain activity oscillatory patterns, we used LTI and calculated a single group-level system dynamics' parameters for each consciousness level -- details in Materials and Methods. Decomposing the estimated system parameters through eigendecomposition revealed the spatiotemporal patterns of the oscillatory modes within the modeled resting-state brain dynamics. 

To investigate how changes in consciousness levels (from awake to deeply sedated) alter the system's spectral profile, we performed $k-$means clustering on the spatial components (eigenvectors) of the eigenmodes across all consciousness states simultaneously. Fig. \ref{Figure_1} presents the eigenvalue distributions and the spatial profiles of the identified centroids for $k=6$ clusters. We further explored the clustering at higher resolutions in the supplementary Figure (SI Fig. \ref{Argand_8}). While there is no unique clustering across scales (SI Fig. \ref{Elbow_method}), we chose $k=6$ to ensure all consciousness states were represented within each cluster. Notably, SI Fig. \ref{Argand_8} demonstrates that similar clusters emerge at higher clustering resolution  ($k=8$).

\begin{figure*}
\centering
\includegraphics[width=1\linewidth ]{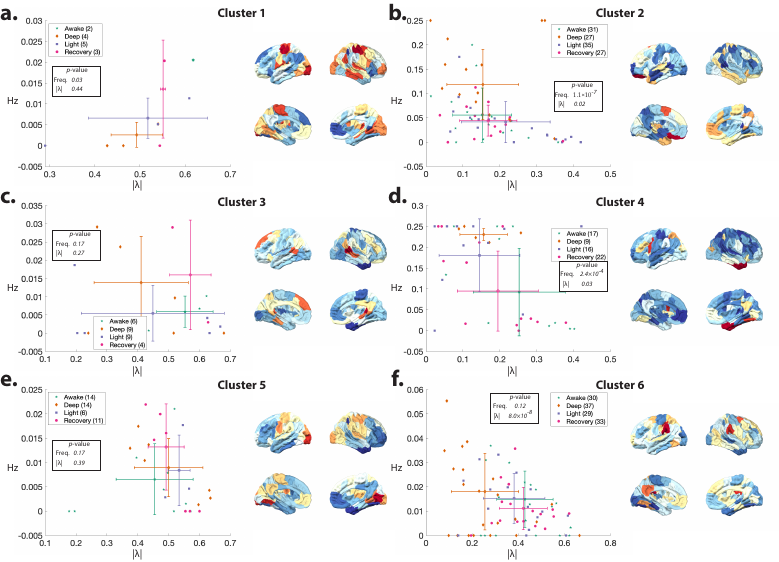}
\caption{\textbf{Sedation-induced changes in the stability and frequency of eigenmodes.} \textbf{a-f} Frequency (Hz) and stability of eigenvalues from the identified $k-$means clusters in Fig. \ref{Figure_1}. The bar plots represent the mean and standard deviation of frequency and stability values for each state of consciousness. The inset legend indicates the number of eigenvalues associated with each consciousness state. The $p$-values indicate the significance of ANOVA tests for frequency and stability across the four states for each cluster. The brain overlay depicts the eigenvector associated with the centroid of the clusters.}
\label{Figure_2}
\end{figure*}

\begin{figure*}
\centering
\includegraphics[width=1\linewidth ]{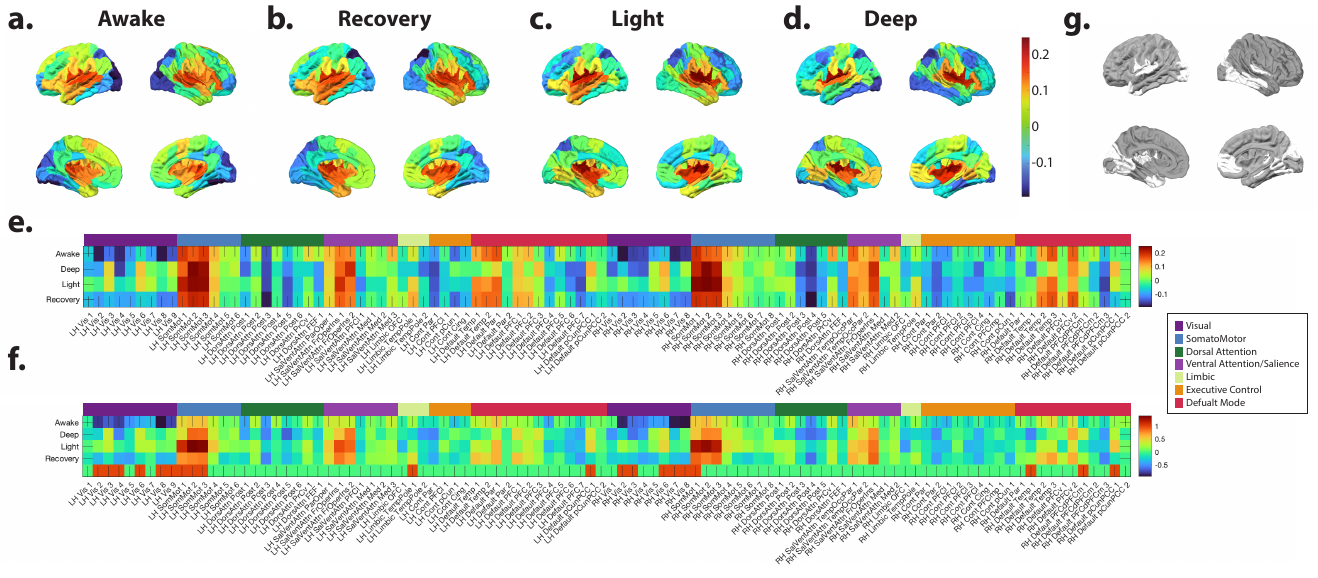}
\caption{\textbf{Principal component analysis (PCA) of the inputs' spatial profiles estimated during auditory stimulation paradigm.} 
\textbf{a-d.} The third principal component (PC3) of the inputs' spatial profiles, estimated during auditory stimulation, shows consistent patterns across different states of consciousness. PC3 captures the activation of the auditory cortex and other active regions. \textbf{e-f.} Average inputs' spatial profiles ($B$) with higher PC3 loading for each subject reveal notable differences despite the overall similarity across different states. The ANOVA test highlights the regions with significant differences ($p<0.05$, FDR corrected for multiple comparisons) across states, denoted by ``1" in the last row and illustrated on the brain overlay in panel \textbf{e}. }
\label{Figure_3}
\end{figure*}

A closer look at the identified clusters revealed distinct changes in the number, frequency, and stability of eigenmodes within each cluster, reflecting the influence of consciousness level. Fig. \ref{Figure_2} illustrates the distribution of frequency and stability of eigenvalues for each cluster. For instance, Cluster 1, comprised of visual, somatomotor, and executive control regions of interest (ROIs), shows a significant reduction of eigenvalue frequencies from awake and recovery toward the light and deep sedation. Whereas Cluster 4, comprised of limbic ROIs, shows the reverse pattern in the frequency. In addition, the stability of eigenvalues of Clusters 3 and~6, comprised of default mode, attention, and executive control ROIs, show a similar pattern of reduction in stability from the awake state toward the sedated state. 

It is worth noticing that the stabilization pattern persists even when using a higher number of clusters ($k=8$). Notably, these higher-resolution clusters share similar spatial profiles with the previously mentioned clusters (SI Fig.~\ref{Angle_stability_8}).  Furthermore, statistical comparisons using ANOVA tests on the frequency and stability distributions across clusters verified the significance of the above-mentioned consciousness-level-dependent changes. These findings demonstrate the effectiveness of this method in capturing identifiable alterations in brain dynamics associated with different consciousness states.

\subsection{Consciousness State-Dependent Co-activation}
\label{sec:ResultsSec2}


\begin{figure*}
\centering
\includegraphics[width=1\linewidth ]{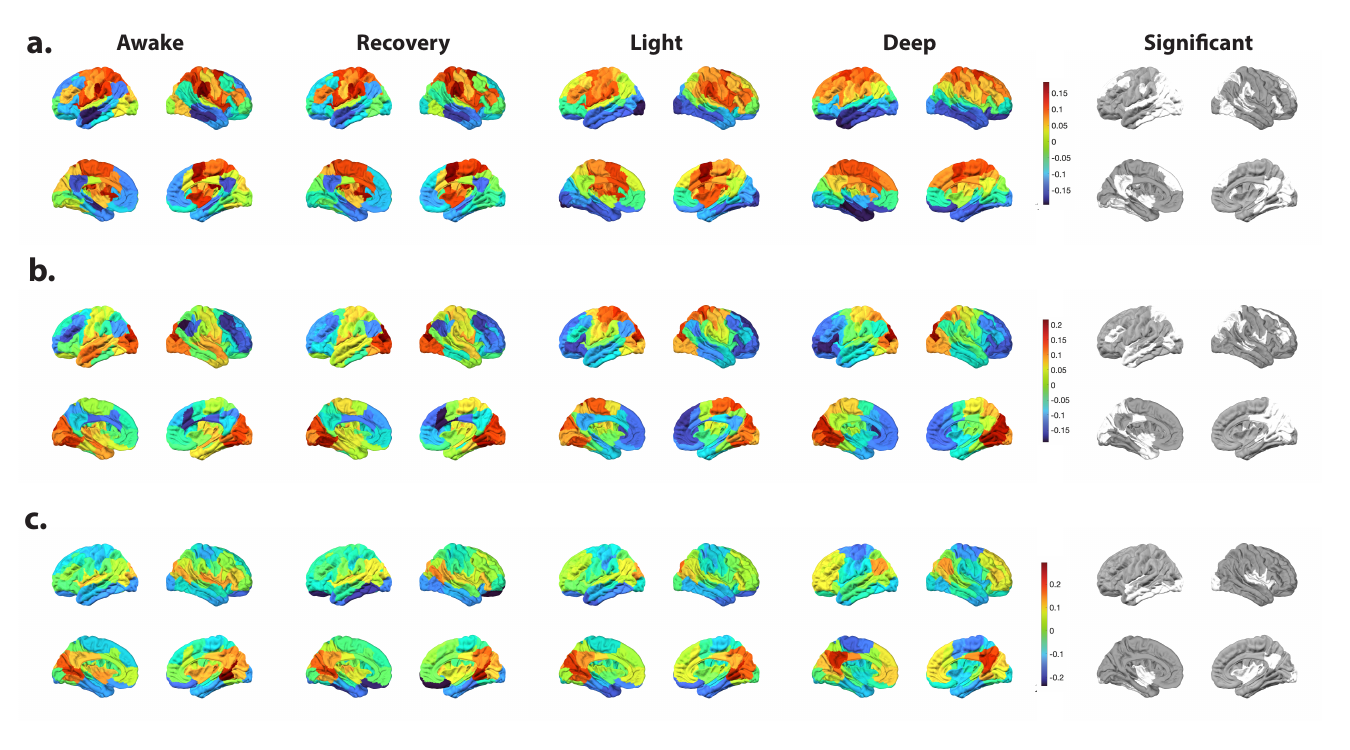}
\caption{\textbf{Principal component analysis (PCA) of the inputs' spatial profiles estimated during auditory stimulation paradigm.} 
The first \textbf{(a)}, second \textbf{(b)}, and fourth \textbf{(c)} principal components of the inputs' spatial profiles, estimated during auditory stimulation. The last panel on the right shows regions with significant variations in the average inputs' spatial profiles ($B$) with the highest loading for each PC across consciousness states, as determined by ANOVA ($p<0.05$, FDR corrected for multiple comparisons).}
\label{Figure_4}
\end{figure*}

\begin{figure*}
\centering
\includegraphics[width=1\linewidth ]{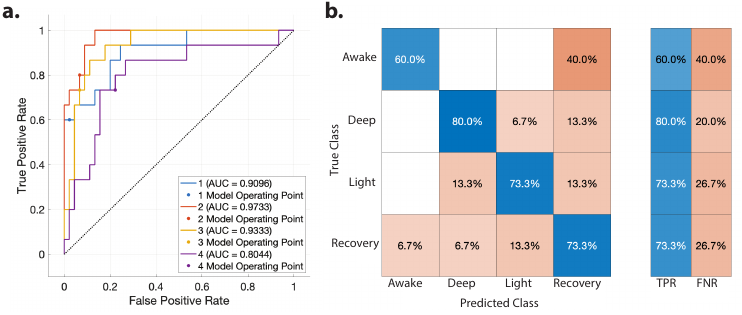}
\caption{\textbf{Classifying  consciousness levels using spatial patterns of external inputs} 
 \textbf{(a)} The Receiver Operating Characteristic (ROC) Curve for classification of different consciousness states with a linear SVM classifier using the vectors of input matrix $B$ associated with PC1 through PC4 from Figs. \ref{Figure_3} and \ref{Figure_4}. \textbf{(b)} Classification's confusion matrix, and true positive and false negative rates of each consciousness level.}
\label{Figure_5}
\end{figure*}


Leveraging our framework with unknown inputs, we investigate how brain responses to auditory stimuli vary across consciousness states. Specifically, it enables the extraction of both spatial ($B$) and temporal ($U$) profiles of external inputs influencing brain activity -- see Materials and Methods for details.  In fact, we hypothesize that the identified input patterns will capture the spread and influence of the stimuli on brain co-activation across consciousness levels. 

First, we considered a lower dimensionality for the inputs ($p=10$) and applied a moderate level of regularization ($\lambda=0.5$) to effectively capture the large-scale driver patterns. This decision stemmed from our analysis, where examining the residuals of the LTI model without external input via principal component analysis (PCA) revealed that approximately $70\%$ of the variance of the residuals could be explained by just ten components (Fig. SI \ref{SI_Figure_4}).

Subsequently, we aggregated the spatial profiles of these inputs (i.e., $B$ matrices) across participants for each consciousness state and employed PCA to discern the key patterns of task-induced activity. Notably, PCA uncovered a consistent PC across consciousness states, exhibiting highly comparable profiles. For example, Fig. \ref{Figure_3}a-d illustrates PC3 based on cortical input profiles at different states -- refer to Materials and Methods for detailed procedures. Notably, this component captures the activation of the auditory cortex in response to the auditory stimulus across all consciousness levels. These results suggest that the activation extends beyond the primary auditory cortex in the temporal lobe during awake and recovery states compared to sedation states. Furthermore, the awake and recovery states exhibit deactivation in the primary visual cortex relative to other states.

To pinpoint cortical regions undergoing the most pronounced changes across different consciousness levels within each participant, we isolated the column vector of $B$  with the highest loading on PC3 for each subject and conducted an ANOVA test. The average identified $B$ vector across participants is depicted in Fig.~\ref{Figure_3}f. These visualizations, along with Fig.~\ref{Figure_3}e, highlight the regions exhibiting significant differences across states ($p < 0.05$, FDR corrected). Noteworthy areas encompass various visual, somatomotor, limbic, and DMN regions.


Additionally, Fig.~\ref{Figure_4} depicts the outcomes of PCA conducted on the cortical $B$ matrices, focusing on the three principal components (PC1, PC2, and PC4). These visualizations also emphasize the regions' significant differences in the average $B$ vectors associated with each PC. Specifically, these PCs correspond to input patterns linked to the attention, somatomotor, and executive control networks (PC1), visual and attention (PC2), visual, executive control, and DMN (PC4). Note that while the estimated inputs' spatial profiles resemble previously identified transient co-activation patterns (CAPs) \citep{huang2020temporal}, their temporal profiles reveal a key difference. Unlike CAPs, these inputs allow for the presence of multiple input patterns at any given time point (SI Fig. \ref{SI_Figure_5}).  

Significantly, ANOVA results, akin to those for the auditory PC3, unveil state-dependent stimulus-induced changes ($p < 0.05$,  FDR corrected). For instance, in PC2, visual inputs predominantly localize to occipital and parietal regions during deep sedation, whereas in the awake state, this pattern extends to temporal regions. Similarly, the PC4 pattern exhibits notable disparities in temporal and posterior cingulate cortex (PCC) regions contingent upon the level of consciousness, with more awake states showing a broader spread in the temporal lobe.

Moreover, we scrutinized the robustness of these findings concerning changes in model hyperparameters (i.e., input dimensions and regularization factor). The results, as illustrated in SI Fig. \ref{SI_Figure_2}, demonstrate that the aforementioned observations are consistently captured even at higher input dimensions ($n=25$ and $\lambda=0.5$) and elevated regularization values ($n=10$ and $\lambda=0.9$). These collective findings underscore the efficacy of our framework in capturing task-induced patterns of large-scale network reconfigurations following alterations in consciousness levels.

Finally, to demonstrate the utility of the spatial input patterns unique to each consciousness level, we employed them to classify consciousness levels. Specifically, we utilized the vectors from the spatial input matrix $B$ associated with the aforementioned PC1-4 across all subjects for consciousness level classification using a linear SVM classifier -- see Materials and Methods for details. Our classification accuracy of 71.7\% underscores the informative nature of the input patterns in tracking consciousness levels. We present the ROC and classification confusion matrix results in Fig. \ref{Figure_5}. 

 \section{Discussion}
\label{sec:Discussion}

 \subsection{Resting-State Dynamics for Disorders of Consciousness Classification and Closed-Loop Control}
\label{sec: Resting-State Dynamics for DOC Classification and Closed-Loop Control}

Resting-state paradigms offer a unique and powerful tool for studying consciousness due to their ability to assess brain activity independent of specific tasks. This advantage allows for direct comparisons across diverse populations, including healthy controls and patients with disorders of consciousness (DOC). Our approach focuses on characterizing the resting brain's spatiotemporal oscillatory patterns, enabling us to track how the level of consciousness modulates the stability and frequency of cortical networks. Our findings provide converging evidence that a hallmark of consciousness loss is the stabilization of oscillatory dynamics. This observation is supported by both electrophysiological data \citep{solovey2015loss,alonso2014dynamical} and computational modeling studies \citep{sanz2021perturbations,piccinini2022data}.

Previous research has demonstrated a link between changes in the blood-oxygen-level-dependent (BOLD) signal frequency and alterations in consciousness state. For example, sleep can be characterized by changes in brain electrical activity, with NREM sleep exhibiting a shift towards low-frequency, high-amplitude EEG patterns. Building on this, recent studies employing simultaneous fMRI and EEG have identified distinct \mbox{low-frequency} and high-frequency BOLD oscillations in the brain during sleep transitions, each with unique spatiotemporal characteristics \citep{song2022fmri}. Our findings complement this work by revealing a significant decrease in the frequency of low-frequency modes encompassing the visual and somatomotor regions and, conversely, an increase in the frequency of high-frequency modes encompassing limbic regions as consciousness levels decline. Moreover, unlike traditional spectral analysis methods like the fast Fourier transform (FFT), which only reveals the temporal frequencies present in a system, our model's identified frequencies correspond to the eigenmodes of the system, capturing both the temporal and spatial nature of the oscillations. 

The observation of unique changes in the spatiotemporal profile of the identified systems' dynamics has significant clinical implications. Firstly, by analyzing the spectral profiles of oscillatory modes, we can potentially extract informative features for patient classification within the multi-dimensional space of consciousness. This information could aid in predicting recovery for patients in difficult-to-classify states. 

More importantly, our framework lays the groundwork for the development of novel therapeutic interventions. By estimating the dynamical system underlying brain activity, we can establish a mathematical objective function for the external control of brain oscillations. Specifically, with knowledge of the current and desired spectral profiles, we can leverage control theory to ``steer" the pathological system toward healthy system dynamics. This could translate into designing targeted feedback stimulation protocols using electrical or transcranial stimulation techniques \citep{medaglia2017brain}. 

Future work should focus on applying this framework to DOC patients. By characterizing individual patient spectral profiles and their evolution during recovery, we can pave the way for closed-loop stimulation protocols. Ultimately, these protocols could be tested to evaluate the model's ability to promote recovery and transition patients from unconsciousness toward wakefulness.

\subsection{Unveiling Consciousness Transitions through Network Reconfiguration Dynamics and Information Integration}
\label{sec:Unveiling Consciousness Transitions through Network Reconfiguration Dynamics and Information Integration}

A significant advantage of our framework lies in its ability to bypass the need for prior knowledge about the external stimulus or the construction of hand-crafted features based on it. For instance, \cite{kandeepan2020modeling} relied on features extracted from the auditory stimulus, including time-domain properties (zero-crossing rate, energy) and frequency-domain properties (spectral centroid, Mel-frequency cepstral coefficients). While such features can identify activity patterns linked to external stimuli, they are limited to capturing low-level stimulus characteristics and cannot inherently capture the stimulus-induced activity on higher-level cognitive processes and activity. Additionally, anticipating relevant low-level features for different modalities can be challenging. As exemplified by \cite{kandeepan2020modeling}, who employed 18 different features to identify potentially relevant ones, this approach can be cumbersome and potentially miss crucial information.

Our framework offers a significant advantage by revealing brain-wide network reconfigurations triggered by external stimuli. This goes beyond simply identifying changes in auditory processing areas, which is expected. Unlike conventional methods that rely on predefined ROIs or predetermined input regressors, our approach can uncover additional stimulus-related activity across various intrinsic brain networks by directly estimating the unknown external inputs driving brain activity.
Therefore, it allows us to capture the combined effects of the stimulus's low-level features and its interaction with higher-level cognitive processes, providing a more comprehensive understanding of stimulus-induced brain dynamics.

 Combining estimated system modes and external inputs sheds light on overlapping changes in network reconfigurations that might be missed by traditional methods like the general linear model (GLM) and FC analyses. For example, our results reveal that auditory stimulus-related input patterns spread beyond the primary auditory cortex in the temporal lobe, aligning with previous findings \citep{kandeepan2020modeling}. These results could indicate reduced complex and elaborated auditory processing in higher-order networks during sedated states. However, we also noted a significant deactivation of early visual areas in PC3 in the awake states, a finding not previously reported by \citep{kandeepan2020modeling}. This suggests that the deactivation profile of visual areas may not be fully captured by the extracted features of the auditory stimulus.

Meanwhile, the PC2 analysis of input patterns reveals that during deep sedation, visual input patterns are confined to the occipital and parietal regions, whereas in wakefulness, they extend into higher-order auditory cortices. These findings indicate dynamic changes in network co-activation across different levels of consciousness. Moreover, the high consciousness level classification accuracy based on the input patterns highlights the consciousness-state-dependence of these activation/deactivation patterns. In wakefulness, auditory stimulation initially activates both primary and higher-order auditory areas, potentially leading to the deactivation of visual areas observed in PC3 patterns. However, the presence of an auditory-visual co-activation pattern in wakefulness (PC2) may also suggest visual processing of auditory information. Conversely, during deep and light sedation, there appears to be more segregated activation of primary sensory systems.

In addition to the auditory task-related co-activation patterns, we demonstrated that the estimated spatial input pattern captures various co-activation/deactivation patterns. In fact, the PCA analysis of these patterns uncovered that these patterns highly resemble the transient, momentary coactivation patterns (CAPs) described in several previous studies \citep{liu2018co,liu2013decomposition,liu2013time,huang2020temporal}. For instance, PC1, PC2, and PC4 are similar to the DAT+/(DMN-), VIS+/(VAT-), DMN+/(DAT-) CAP identified in \citep{huang2020temporal}. However, the primary distinction between CAPs and the identified inputs in the LTI system is that, unlike CAPs, multiple input patterns can coexist simultaneously at any given time point. 

Overall, our framework offers a more nuanced understanding of how information integrates across modalities and intrinsic brain systems during different consciousness states. This has significant clinical implications, particularly for DOC patients. By combining complex naturalistic stimuli with the input patterns identified by our model, we can potentially develop more objective and reliable methods for assessing awareness in DOC patients. This approach could involve examining how information is processed and shared across sensory and higher-level associative brain networks. This could lead to improved diagnosis, prognosis, and, ultimately, the development of targeted interventions for DOC patients.
 
\subsection{Limitations and Future Directions}
\label{sec:Limitations and Future Directions}

This study has several limitations that motivate future research directions. First, we utilized a publicly available dataset with a limited scan duration per subject. While we mitigated this by estimating a single \mbox{group-level} system matrix to capture slow brain oscillations, ideally, future studies should employ longer resting-state scans (15-30 minutes) for each participant to enable subject-specific system models, potentially leading to more accurate predictions.

Second, ventral brain regions generally exhibit lower signal-to-noise ratios (SNR) \citep{rua2018improving}. To address this, we excluded participants with missing ROI data and lowered the threshold for some brain voxels to include these regions. This approach might have introduced additional noise, and future work should utilize datasets specifically designed to enhance SNR for a more reliable assessment of these regions.

Finally, our framework is built on the assumption of time-invariant brain network oscillations. This assumption is partly supported by the dependence of cortical dynamics on their static structural scaffold and by theoretical and modeling work. However, the question of whether a time-varying system would offer a more accurate representation of large-scale brain dynamics and improved predictive power remains open. While it's challenging to distinguish between a \mbox{time-invariant} system with time-varying inputs and a truly time-varying system, future studies could compare our findings against time-varying system models. This could potentially lead to a deeper understanding of large-scale brain dynamics and their applications in forecasting recovery trajectories for DOC patients.

\section{Conclusions}
\label{sec: Conclusions}

This study sheds light on the neural underpinnings of consciousness by analyzing the interplay between spatiotemporal oscillatory patterns and their external drivers within large-scale brain networks. We demonstrate a critical link between consciousness levels and the dynamics of these networks, with a shift towards stabilized oscillations characterizing unconsciousness. Importantly, our framework offers a unique approach to studying consciousness, bypassing the need for predefined features and revealing consciousness-level-dependent brain-wide reconfigurations of external drivers of brain dynamics. Significantly impacting clinical care, our research can guide the development of objective assessment tools and targeted interventions for disorders of consciousness.

\section{Materials and Methods}
\label{sec: Materials and Methods}

\subsection{Linear time-invariant (LTI) dynamical systems with external inputs}
\addcontentsline{toc}{subsection}{Linear time-invariant (LTI) dynamical systems with external inputs.}

Each region of interest $i$  provides a time series denoted by $x_i[k]$ at sampling point $k=0,\ldots, T$. We consider a total of $n =100$ cortical ROIs. These signals are collectively represented by the vector $x[k]=[x_1[k] \ \ldots \ x_n[k]]^\intercal$, with $k=0,\ldots, T$, referred to as the state of the system, describing the BOLD signal's evolution across regions. The system's state evolves primarily due to (\emph{i}) cross-dependencies among signals from different regions and (\emph{ii}) external inputs, which may be excitation noise or unaccounted extrinsic stimuli.

To model the system's state evolution, we propose

\begin{equation}
x[k+1]=Ax[k]+ Bu[k]+\omega_k, \quad k=0,\ldots, T,
\label{eq:systemDyn}
\end{equation}

\noindent where $A\in\mathbb{R}^{n \times n}$ represents the coupling dynamics, $B\in\mathbb{R}^{n\times p}$ is the input matrix describing the influence of inputs $u[k]\in\mathbb{R}^{p\times 1}$ on state evolution, and $\omega_k\in\mathbb{R}^n$ is internal dynamics noise at sampling point $k$. Notably, $\{x[k]\}_{k=0}^T$ denotes BOLD signals across ROIs, being the only known information. However, the underlying neural activity state remains unknown due to the absence of the hemodynamic response function in our model. Hence, the input in the model reflects external drivers of regional BOLD, indirectly capturing neural activity. To determine the system parameters ~\eqref{eq:systemDyn} ($A$, $B$, $\{u[k]\}_{k=0}^T$), we minimize the distance between the system's state $x[k]$ and the estimated state $\hat x[k]$ driven by unknowns, yielding the optimization problem:

\begin{eqnarray*}
\{\hat x[k]\}_{k=0}^T \in \arg\min_{z[0],\ldots,z[T]}& \quad  \|z[k]-x[k]\|_2^2 \\
\text{ s.t.} & z[k+1]=Az[k]+ Bu[k].
\end{eqnarray*} 


This problem is more complex than standard least squares due to unknown system parameters \citep{ljung1999system}. Therefore, following \cite{PequitoC35}, we undertake the following steps: (i) setting $z[0]=x[0]$ and $\{u[k]\}_{k=0}^T$ to zero to approximate $A$; (ii) assuming $A$ from step (i), providing a sparse low-rank structure to $B$ to approximate $z[0]$ and $\{u[k]\}_{k=0}^T$, yielding subsequent $z[0],\ldots,z[T]$, $\{\hat{x}[k]\}_{k=0}^T$; and (iii) assuming $\{z[k]\}_{k=0}^T$ and $\{u[k]\}_{k=0}^T$ as approximated, obtaining an approximation for $B$. Steps (ii) and (iii) are performed iteratively until the parameter estimates converge (typically within a few iterations). To prevent the external inputs from solely capturing all the information, we penalize their use in the optimization objective function. This is achieved by adding a regularization factor (i.e., sparsity term, $\|z[k]-x[k]\|_2^2 + \lambda \|u\|_1+ \lambda \|B\|_1^2$ with weight $\lambda>0$) that discourages overly complex input patterns. For detailed algorithmic procedures, refer to \ref{sec: SI1}.

We demonstrated in our previous work that unaccounted external inputs result in errors in the estimation of system matrix $A$ \citep{ashourvan2022external}. Therefore, in a modified version of this algorithm, in step (i), we estimate $A$ from $ x[k]$ measured during resting-state scans (i.e., an extended period without task-related external stimulation). Next, we iteratively repeat steps (ii) and (iii) as detailed above.

The shorter resting-state scans in the \cite{kandeepan2020modeling} dataset posed limitations on individual-level parameter estimation. To overcome this constraint, rather than computing separate system matrices based on the LTI model for each subject, we derived a single group-level system parameter for each consciousness level. This was achieved through simultaneous minimization of the least squared error across all participants using unconstrained nonlinear optimization employing a quasi-Newton algorithm \citep{broyden1970convergence,shanno1970conditioning}.

For the mathematical description of the cost function for the explained optimization problem of estimating a single $A$ matrix of an autonomous LTI system without external inputs using the least squared error across all subjects simultaneously, we can represent it as follows:
Given a set of observations $x_i$ for $N$ subjects over $T$ time points and a model prediction $\hat{x}_i$ based on the LTI system with a system matrix $A$, the cost function can be defined as the sum of squared errors across all subjects:


\begin{eqnarray*}
J(A) = \sum_{i=1}^{N} \sum_{k=1}^{T} (x_i[k] - \hat{x}_i[k])^2 \\
\text{s.t.} \quad \hat{x}_i[k+1] = A\hat{x}_i[k] + Bu[k],
\end{eqnarray*}

\noindent where $x_i[k]$ represents the observed data for subject $i$ at time $k$, and $\hat{x}_i[k]$, is the model prediction based on the LTI system with system matrix $A$.

The optimization problem is then to find the system matrix $A$ that minimizes this cost function: $\min_{A} J(A)$. This optimization is performed using iterative algorithms such as the quasi-Newton method mentioned earlier, which iteratively updates the estimate of $A$ until convergence to a minimum of the cost function is achieved.

Since we did not know the true dimensionality of the external inputs, we approximated the dimensions of the input matrix $B$ by performing principal component analysis on the residuals of the models. As seen in \mbox{SI Fig. \ref{SI_Figure_4}}, the first 10 and 25  PCs capture more than $\approx 70\%$ and $\approx 90\%$ of variance in the average residuals across all tasks, respectively.  


In addition, we demonstrate that we identify external input patterns during the auditory stimulation task and consciousness levels similarly at both low and high-dimensional input matrices (SI Fig.\ref{Auditory_10_25}). Therefore, we select $p = 10$ for input matrix $B$ to estimate the inputs in the main manuscript to capture the large-scale cortical input patterns.

\subsection{Eigenmode Decomposition for Brain Dynamics Analysis}
\addcontentsline{toc}{subsection}{Eigenmode Decomposition for Brain Dynamics Analysis}

Our analysis leverages the concept of eigenmode decomposition to understand the dynamic behavior of the brain's BOLD signal.  Given an LTI description of the system dynamics, we can decompose the evolution of this system into its eigenmodes.

\subsubsection{Eigenmodes and their Properties}
\addcontentsline{toc}{subsubsection}{Eigenmodes and their Properties}

An eigenmode is characterized by an \mbox{eigenvalue-eigenvector} pair $(\lambda_i, v_i)$.  The system dynamics satisfy the equation $Av_i = \lambda_i v_i$, where $A$ is the system matrix, $v_i$ is the eigenvector corresponding to the eigenvalue $\lambda_i$.  Each eigenmode describes the oscillatory behavior of the system along a specific direction defined by the eigenvector $v_i$.

The eigenvalue $\lambda_i$ itself holds valuable information about the dynamics in that direction:

\begin{itemize}

\item \textbf{Frequency}: Represented in polar coordinates by $(\theta_i, |\lambda_i|)$, the frequency of the oscillation associated with the eigenmode is
\[
f_i = \frac{\theta_i}{2\pi} \delta t,
\]
where $\delta t$ is the sampling frequency of the data.

\item \textbf{Stability} (Damping Rate): The Stability or time scale, which reflects how quickly the oscillation decays or grows over time, is captured by
\[
\rho_i = \frac{\log(|\lambda_i|)}{\delta t}.
\]

\end{itemize}

The interpretation of the time scale depends on the magnitude of the eigenvalue:

\begin{itemize}

   \item \emph{Damping (Stable)}: If $|\lambda_i| < 1$, the magnitude of the oscillation along that direction decays to zero over time, indicating a stable process;
   \item \emph{Growing (Unstable)}: If $|\lambda_i| > 1$, the magnitude of the oscillation grows without bound, indicating an unstable process, and
   \item \emph{Meta-Stable}: If $|\lambda_i| \approx 1$, the process oscillates between periods of stability and instability, exhibiting a meta-stable behavior.
\end{itemize}

\subsubsection{From Eigenvectors to Spatial Contributions}
\addcontentsline{toc}{subsubsection}{Fr  Eigenvectors to Spatial Contributions}

The eigenvector matrix, denoted by $V = [v_1, ..., v_n]$, contains all the eigenvectors as columns.  We can express the system dynamics in terms of these eigenvectors using a change of variable:
\[
z[k] = V^* x[k]
\]
Here, $V^*$ is the conjugate transpose of $V$, $x[k]$ is the original state vector of the system at time step $k$, and $z[k]$ is the transformed state vector.  The $i^{th}$ component of $z[k]$, denoted by $z_i[k]$, represents a weighted combination of the original state variables based on the $i^{th}$ eigenvector, $v_i$.  Therefore, $z_i[k]$ captures the specific spatial contributions of the different ROIs to the overall brain activity at the spatiotemporal frequency characterized by the eigenvalue $\lambda_i$.

By analyzing the eigenmodes of the system dynamics, we can extract key information about the brain's BOLD signal evolution.  The eigenvalues reveal the timescales of the underlying processes, while the eigenvectors describe the spatial contributions of different ROIs. Together, this decomposition provides a comprehensive understanding of the spatiotemporal dynamics of brain activity.

\subsection{Dataset and Preprocessing}
\addcontentsline{toc}{subsection}{Dataset and Preprocessing} 

We used a publicly available dataset from \cite{kandeepan2020modeling}, which was accessed from Openneuro.org \citep{Kandeepands003171}. The dataset protocol included 17 healthy participants (4 women; mean age = 24 $\pm  5$) with no history of neurological disorders. Participants completed fMRI at four levels of sedation (awake, mild sedation, deep sedation, and recovery) during resting-state scans as well as while listening to a 5-minute audio recording from the movie ``Taken". Functional echo-planar images (EPI) were acquired at a matrix size of 64 x 64 with a spatial resolution of 3 mm isotropic voxels. Images contain 33 slices with a 25\% inter-slice gap with a repetition time (TR) of 2000 ms and time echo (TE) of 30 ms. Audio task and resting-state scans had 155 and 256 samples, respectively. An anatomical scan was also obtained using a T1-weighted 3D MPRAGE (magnetization prepared rapid gradient echo) sequence. Anatomical image acquired at a matrix size of 240 x 256 x 192 with a spatial resolution of 1 mm isotropic voxels and 4250 ms TE. 

The dataset obtained from Openneuro was preprocessed through the fMRIprep preprocessing pipeline. T1w images in the data were used to create a reference T1w to correct for intensity non-uniformity with N4BiasFieldCorrection (ANTs) \citep{tustison2010n4itk}. The reference was then skull-stripped using a NiPype implementation of the antsBrainExtraction.sh (ANTs) workflow tool using the OASIS brain extraction template \citep{marcus2007open} as a target. Brain surfaces were reconstructed from the reference T1w image using the FreeSurfer tool recon-all \citep{dale1999cortical}. Brain tissue segmentation of gray matter, white matter, and cerebrospinal fluid was computed from the reference T1w image using FSL's FAST \citep{zhang2001segmentation}, spatial normalization to the ICBM Nonlinear Asymmetrical template (MNI152NLin2009cAsym) was performed using antsRegistration (ANTs). 

For BOLD images (EPI), a reference image was created from the median of motion-corrected BOLD images. Head motion is estimated using FSL's mcflirt \citep{jenkinson2002improved}. The BOLD runs were then slice-timing corrected using AFNI's 3dTshift \citep{cox1997software} and underwent susceptibility distortion correction (SDC). These files are then aligned using the gray/white-matter boundary and resampled to MNI152NLin2009cAsym and fsaverage \citep{zhang2001segmentation} (Freesurfer) template space.

The preprocessed BOLD images underwent further processing using the eXtensible Connectivity Pipeline-DCAN (XCP-D) postprocessing pipeline \citep{mehta2023xcp}. Postprocessing denoising of the data included confound regression of nuisance regressors using the 36P strategy configuration, which includes six realignment motion parameters, white matter, CSF, and global signal parameters. To retain as much data in the final output, temporal censoring and data filtering were disabled. For the output final step, minimum coverage was set to 0.01. We excluded 2 participants with any ROIs that did not meet this criteria. Voxel-wise time series were extracted from the denoised BOLD images and parcellated to the combined 4S atlas \citep{cieslak2021qsiprep}. From this combined atlas, we utilized Schaefer cortical atlas \citep{schaefer2017local} at 100-region resolution for analysis.

\subsection{Statistics}
\addcontentsline{toc}{subsection}{Statistics}

\subsubsection{Identifying Shared Eigenmode Profiles:}

We employed k-means clustering to identify groups of eigenmodes exhibiting similar spatial profiles across all consciousness states.  Calinski-Harabasz \citep{Calinski1974}, Davies-Bouldin\citep{Davies1979}, and Silhouette\citep{Rousseeuw1987} criteria were used to assess the optimal clustering resolution (see \ref{sec:SI3} for details). However, these criteria yielded inconsistent results, suggesting that the eigenvector clusters lack a well-defined optimal number of communities at a specific topological scale.  The elbow method, which analyzes explained variance versus the number of clusters, further confirmed the absence of a clear optimal clustering resolution (SI Fig. \ref{Elbow_method}).

Therefore, we opted for a data-driven approach, selecting a number of clusters ($k=6$) that ensures all consciousness states are represented within each cluster.  Subsequently, we examined potential differences in the stability and frequency of eigenvalues across these clusters.  This analysis involved evaluating the frequency and radius (i.e., stability) of eigenvalues within each cluster for all four consciousness states using Analysis of Variance (ANOVA) with a significance threshold of $p < 0.05$.

To identify estimated inputs reflecting changes corresponding to different consciousness levels, we conducted a principal component analysis (PCA) on the concatenated spatial profiles ($B$) of all estimated inputs across all subjects for each consciousness state. Subsequently, we determined a single input with the highest absolute principal component (PC) loading for each component. Inputs with negative PC loadings were multiplied by $-1$.  
Next, for each ROI, we performed an ANOVA to assess the significance of differences across the means among the four consciousness levels ($p < 0.05$). We corrected the calculated test statistics for multiple comparisons across all ROIs using the false discovery rate (FDR) method \citep{benjamini1995controlling}.

\subsubsection{Consciousness state calssification:}

We implemented a Linear Support Vector Machine (SVM) classifier to predict the consciousness state based on the vectors from the spatial
input matrix $B$ associated with the first four leading PCs concatenated $B$ matrices across all subjects. The training process involved several preprocessing steps to ensure the data was suitable for modeling. Initially, we applied PCA to feature sets to the numeric predictor variables to reduce the dimensionality of the data. We retained enough principal components to explain 95\% of the variance in the predictor data. This reduction helped enhance computational efficiency and improve the model's performance by focusing on the most significant features.

We trained the SVM classifier with the preprocessed data, which involved defining a prediction function to enable future predictions on new data. The performance of the trained classifier was evaluated using five-fold cross-validation, yielding a validation accuracy that reflects the model's predictive capability. The trained classifier and its validation accuracy were then outputted for further use and assessment.

The ROC curves for each class label were generated by computing the true positive rate and false positive rate for various threshold settings. By plotting these rates, we create ROC curves that visually represent the model's performance in distinguishing between classes. The area under each ROC curve (AUC) indicates the model's ability to correctly classify instances of each class, with higher AUC values signifying better performance.


 \section*{Data Availability}
We used a publicly available dataset from \cite{kandeepan2020modeling}, which was accessed from Openneuro.org \citep{Kandeepands003171}. 

\section*{Code Availability}
The custom scripts are available at https://github.com/aashourv/LTI\_Group/.

\section*{Acknowledgements}    
 This work was supported by the startup funding from the Department of Psychology at the University of Kansas provided to A.A." 

\section*{Author Contributions}
J.B.: Conceptualization of this study, Data Preprocessing, Methodology, Data Analysis, Writing.
S.P.: Conceptualization of this study, Methodology, Writing. 
P.B.: Conceptualization of this study, Methodology, Writing. 
A.A.: Conceptualization of this study, Methodology, Data Analysis, Writing. 

\bibliographystyle{elsarticle-harv} 
\bibliography{cas-refs}

\clearpage
 \newpage
 
\appendix
\section{Joint-estimation algorithm.}
\label{sec: SI1}

In what follows, we seek to determine the dynamics and input matrices, as well as the input sequence of the following dynamical system,

\begin{equation}\label{eqn:system}
x[k+1]=Ax[k]+Bu[k],
\end{equation}

\noindent where $x[k]\in\mathbb{R}^n$ denotes the state, $u[k]\in\mathbb{R}^p$ denotes the input, $A\in\mathbb{R}^{n\times n}$ the dynamics matrix, and $B\in \mathbb{R}^{n\times p}$ the input matrix. Specifically, by considering a sequence of data $\{z[k]\}_{k=0}^{N-1}$, we seek to find an approximation of the parameters $(A,B,\{u[k]\}_{k=0}^{N-1})$ denoted by $(\tilde A,\tilde B, \{\tilde u[k]\}_{k=0}^{N-1})$, respectively. Next, notice that given the dynamical system~\eqref{eqn:system}, we obtain

\begin{equation}
z[k]=Az[k-1]+Bu[k-1]+\varepsilon_k,
\label{eqn:zError}
\end{equation}

\noindent where the error can be captured by $\varepsilon_k\sim\mathcal{N}(0,\Sigma)$. Subsequently, in a least-squares minimization sense, we seek to minimize the following

\begin{equation}
\min_{(\tilde A, \tilde B,\{\tilde u[k]\}_{k=0}^{N-1})} \sum_{k=1}^N \|z[k]-\left(\tilde A z[k-1]+\tilde B\tilde u[k-1]\right)\|_2^2,
\label{eqn:LS}
\end{equation}

Additionally, since both the input matrix and sequence of inputs are unknown, the problem is not well posed, which forces us to consider possible feasibility constraints (e.g., $\|\tilde u[k]\|\le 1$) and the objective should be changed to account for a regularization term (e.g., the 1-norm) that penalize the number of non-zero entries, which lead to the following problem.

\begin{equation}
\min_{(\tilde A, \tilde B,\{\tilde u[k]\}_{k=0}^{N-1})} \sum_{k=1}^N \|z[k]-\left(\tilde A z[k-1]+\tilde B\tilde u[k-1]\right)\|_2^2+\lambda \|\tilde u[k-1]\|_1^2,
\label{eqn:LSreg}
\end{equation}

\noindent where $\lambda\in\mathbb{R}_0^{+}$ is the regularization term that weights the tradeoffs between the approximation error and the number of nonzero entries. To determine the unknown parameters  $(\tilde A,\tilde B, \{\tilde u[k]\}_{k=0}^{N-1})$ that is the solution to the above problem, we will consider an algorithm that borrows ideas from the Expectation-Maximization algorithm, which details can be found in \cite{Gupta2019, PequitoC35}. Briefly, we seek to determine a converging sequence of unknown parameters $(\tilde A^{(l)},\tilde B^{(l)}, \{\tilde u^{(l)}[k]\}_{k=0}^{N-1})$, which initial parameters (i.e., with $l=0$) are set as follows: (\emph{i}) $A^{(0)}$ is the solution to~\eqref{eqn:LS} with input matrix and inputs set to zero; and (ii) $\tilde B^{(0)} $ is initialized at random which entries are obtained from standard Gaussians. Next, for a predefined value of $\lambda$ in~\eqref{eqn:LSreg}, we find a solution to it by proceeding sequentially as follows (for each iteration $l=1,2,\ldots$):

noindent (\emph{i}) determine $\{\tilde u^{(l)}[k]\}_{k=0}^{N-1}$, given the values $(\tilde A^{(l-1)}, \tilde B^{(l-1)})$
 
\begin{equation}
\min_{\{\tilde u^{(l)}[k]\}_{k=0}^{N-1}} \sum_{k=1}^N \|z[k]-\left(\tilde A^{(l-1)} z[k-1]+\tilde B^{(l-1)}\tilde u^{(l)}[k-1]\right)\|_2^2+\lambda \|\tilde u^{(l)}[k-1]\|_1^2,
\label{eqn:LSreg1}
\end{equation}

\noindent (\emph{ii})~determine $\tilde A^{(l)}$ given the values $(\tilde B^{(l-1)}, \{\tilde u^{(l)}[k]\}_{k=0}^{N-1})$, which is the solution to the following problem
\begin{equation}
\min_{(\tilde A^{(l)})} \sum_{k=1}^N \|z[k]-\left(\tilde A^{(l)} z[k-1]+\tilde B^{(l-1)}\tilde u^{(l)}[k-1]\right)\|_2^2+\lambda \|\tilde u^{(l)}[k-1]\|_1^2,
\label{eqn:LSreg2}
\end{equation}

 \noindent (\emph{iii}) determine $\tilde B^{(l)}$ given the values $(\tilde A^{(l)}, \{\tilde u^{(l)}[k]\}_{k=0}^{N-1})$, which is the solution to the following problem
\begin{equation}
\min_{(\tilde B^{(l)})} \sum_{k=1}^N \|z[k]-\left(\tilde A^{(l)} z[k-1]+\tilde B^{(l)}\tilde u^{(l)}[k-1]\right)\|_2^2+\lambda \|\tilde u^{(l)}[k-1]\|_1^2,
\label{eqn:LSreg3}
\end{equation}

\noindent and (\emph{iv})~determine $\tilde A^{(l)}$ given the values $(\tilde B^{(l)}, \{\tilde u^{(l)}[k]\}_{k=0}^{N-1})$, which is the solution to the following problem
\begin{equation}
\min_{(\tilde A^{(l)})} \sum_{k=1}^N \|z[k]-\left(\tilde A^{(l)} z[k-1]+\tilde B^{(l)}\tilde u^{(l)}[k-1]\right)\|_2^2+\lambda \|\tilde u^{(l)}[k-1]\|_1^2.
\label{eqn:LSreg4}
\end{equation}

Lastly, it is important to notice that several criteria can be adopted to stop the sequential procedure in the algorithm. In particular, we have considered the total variation between iterations on the inputs to be below a threshold. 

 \clearpage
 \newpage

\section{Determining the optimal number of eigenvector clusters.}
\label{sec:SI3}

\addcontentsline{toc}{section}{SI4. Determining the optimal number of eigenvector clusters}

To find the optimal number of eigenvector clusters, we used the following methods.

\subsection*{Elbow Method}
K-means clustering aims to minimize the within-cluster variance or the within-cluster sum of squared errors (SSE). Ideally, this error should be as small as possible. However, increasing the number of clusters always reduces the SSE. The Elbow method involves plotting SSE as a function of the number of clusters and identifying a point (the "elbow") where adding another cluster does not significantly reduce the SSE. This elbow typically indicates the optimal number of clusters, although it may not always be easily identifiable, especially in less clustered data.

\subsection*{Calinski-Harabasz Criterion}
We used the Calinski-Harabasz criterion \citep{Calinski1974}, also known as the variance ratio criterion, defined as:

\begin{equation}
VRC = \frac{SS_{B}}{SS_{W}} \times \frac{(N-k)}{(k-1)},
\end{equation}

where $k$ is the number of clusters, $N$ is the number of observations, and $SS_{B}$ and $SS_{W}$ are the total between-cluster and within-cluster variance, respectively. $SS_{B}$ and $SS_{W}$ are defined as:

\begin{equation}
SS_{B} = \sum_{i=1}^{k} n_{i} \left\| m_{i} - m \right\|^{2},
\end{equation}

\begin{equation}
SS_{W} = \sum_{i=1}^{k} \sum_{x \in c_{i}} \left\| x - m_{i} \right\|^{2},
\end{equation}

where $n_{i}$ is the number of points in the $i$th cluster, $m_{i}$ is the centroid of the $i$th cluster, $m$ is the overall mean of the data, and $\left\| \cdot \right\|$ denotes the Euclidean distance. The optimal number of clusters maximizes the variance ratio criterion, indicating well-defined clusters with high between-cluster variance and low within-cluster variance.

\subsection*{Davies-Bouldin Criterion}
The Davies-Bouldin criterion \citep{Davies1979} measures the ratio of within-cluster to between-cluster distances and is defined as:

\begin{equation}
DB = \frac{1}{k} \sum_{i=1}^{k} \max_{j \neq i} D_{i,j},
\end{equation}

where $D_{i,j}$ is the ratio of within-cluster to between-cluster distance for clusters $i$ and $j$, defined as:

\begin{equation}
D_{i,j} = \frac{\bar{d_{i}} + \bar{d_{j}}}{d_{i,j}},
\end{equation}

where $\bar{d_{i}}$ and $\bar{d_{j}}$ are the average distances of data points in clusters $i$ and $j$ to their respective centroids, and $d_{i,j}$ is the Euclidean distance between the centroids of clusters $i$ and $j$. The optimal number of clusters minimizes the Davies-Bouldin index, representing the best within-to-between cluster distance ratio.

\subsection*{Silhouette Criterion}
The Silhouette criterion \citep{Rousseeuw1987} measures how similar each point is to its own cluster compared to other clusters and is defined as:

\begin{equation}
S_{i} = \frac{b_{i} - a_{i}}{\max(a_{i}, b_{i})},
\end{equation}

where $a_{i}$ is the average distance between the $i^{th}$ data point and other points in its cluster, and $b_{i}$ is the minimum average distance between the $i^{th}$ data point and points in different clusters. A high Silhouette score (ranging between 1 and -1) indicates that the data point is well-clustered within its own cluster and poorly matches data points from other clusters. Conversely, many data points with zero or negative Silhouette values suggest the presence of too few or too many clusters.

 \clearpage
 \newpage

\section{Supplementary Figures.}
\label{sec:SI4}
\setcounter{figure}{0}

\begin{figure*}
\centering
\includegraphics[width=1 \linewidth ]{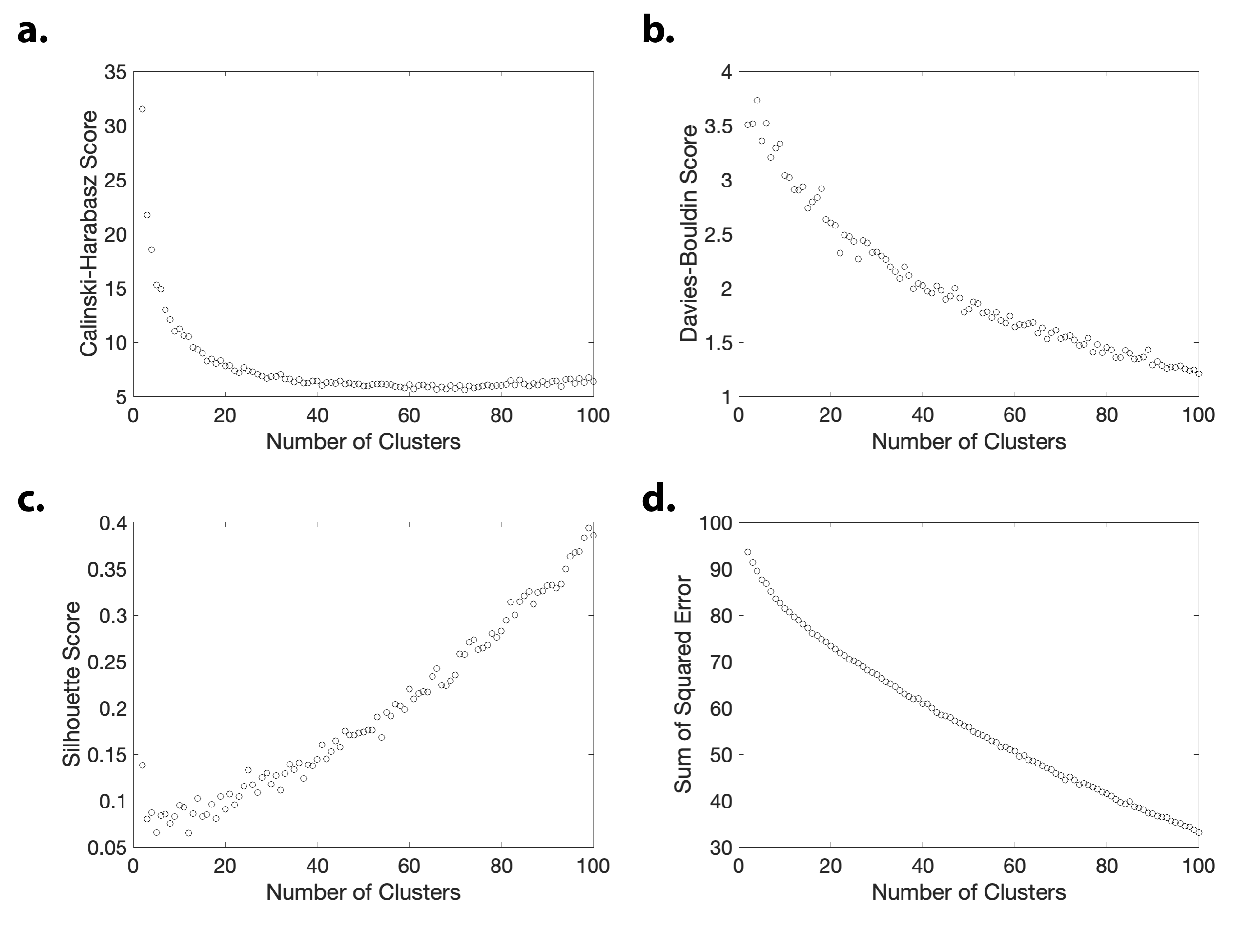}
\caption{\textbf{Determining the Number of Eigenvector Clusters.} To evaluate the optimal number of clusters $(k)$ present in the eigenvectors, we used the \emph{\textbf{(A)}} Calinski-Harabasz criterion (ratio of within-cluster to between-cluster dispersion) \citep{Calinski1974}, \emph{\textbf{(B)}} Davies-Bouldin criterion (ratio of within-cluster to between-cluster distances) \citep{Davies1979}, and \emph{\textbf{(C)}} Silhouette criterion (measure of similarity of each data point to other points in its cluster compared to points in other clusters) \citep{Rousseeuw1987}. Based on these three criteria, the optimal number of clusters in the data is $k = 2, 100,$ and 99, respectively, as shown in panels \textbf{A-C}. \emph{\textbf{(D)}} The plot represents the sum of squared errors (distances to cluster centers). Due to the smoothness of the curve, a single $k$ corresponding to the \emph{`elbow'} of the curve in panel \emph{\textbf{D}} cannot be identified. Together, these results suggest that the data may not contain clearly defined clusters for $k > 2$. Therefore, to examine the organization of eigenvectors at higher resolutions in the main manuscript, we choose $k = 6$, which roughly corresponds to the elbow of the curve.}
\label{Elbow_method}
\end{figure*}

\begin{figure*}
\centering
\includegraphics[width=1\linewidth ]{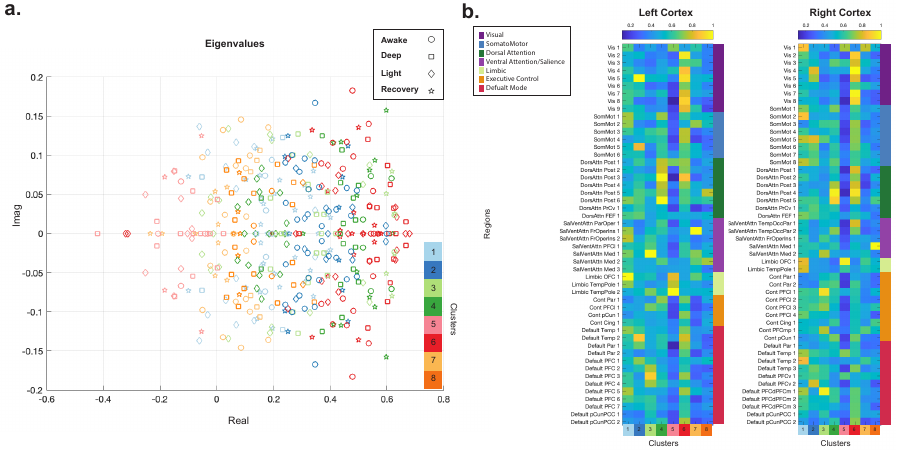}
\caption{\textbf{Spectral profiles of brain's oscillatory modes at different levels of consciousness.} \textbf{a.} Distribution of eigenvalues from \mbox{resting-state} scans across varying states of consciousness. The identified eigenvector clusters ($k=8$) are represented with distinct colors, and different consciousness states (awake, light sedation, deep sedation, and recovery) are denoted by different markers. \textbf{b.} Identified centroid of eigenvector clusters (color-coded column-wise) associated with the eigenvalue from panel \textbf{a}.}
\label{Argand_8}
\end{figure*}

\begin{figure*}
\centering
\includegraphics[width=1\linewidth ]{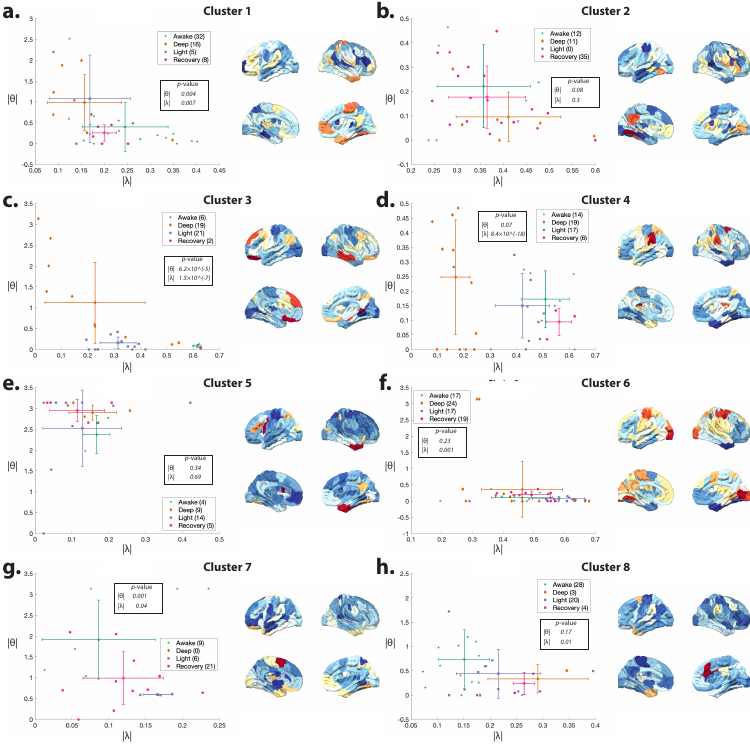}
\caption{\textbf{Sedation-induced changes in the stability and angle of eigenmodes.} \textbf{a-h} Angle and radius (stability) of eigenvalues from the identified $k-$means clusters in SI Fig. \ref{Argand_8}. The bar plots represent the mean and standard deviation of Angle and stability values for each state of consciousness. The inset legend indicates the number of eigenvalues associated with each consciousness state. $p$-values indicate the significance of ANOVA tests for Angle and stability across the four states for each cluster. The brain overlay illustrates the centroids of the clusters.}
\label{Angle_stability_8}
\end{figure*}

\begin{figure*}
\centering
\includegraphics[width=1\linewidth ]{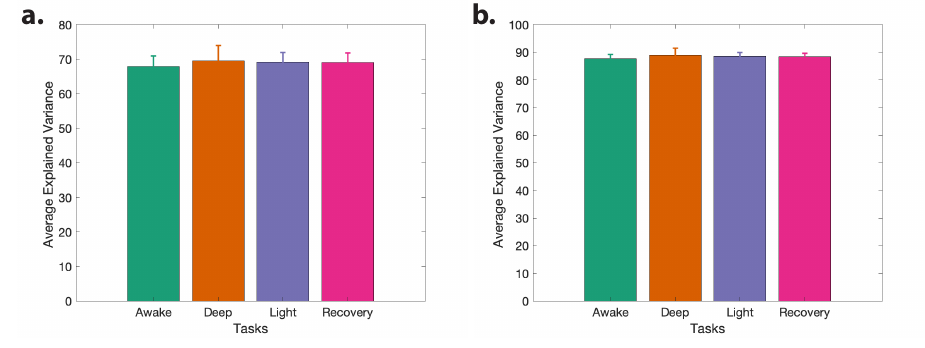}
\caption{\textbf{Principal Component Analysis (PCA) of the LTI model's residuals during auditory stimulation paradigm.} The mean and standard deviation (error bars) of the explained variance using the first 10 \textbf{(a)} and 25 \textbf{(b)} PCs calculated from the residuals of the LTI model using auditory stimulation scans across different consciousness levels. The average mean explained variance across all consciousness levels is 68.9\% and 88.4\% in panels \textbf{a} and \textbf{b}, respectively.}
\label{SI_Figure_4}
\end{figure*}

\begin{figure*}
\centering
\includegraphics[width=1\linewidth ]{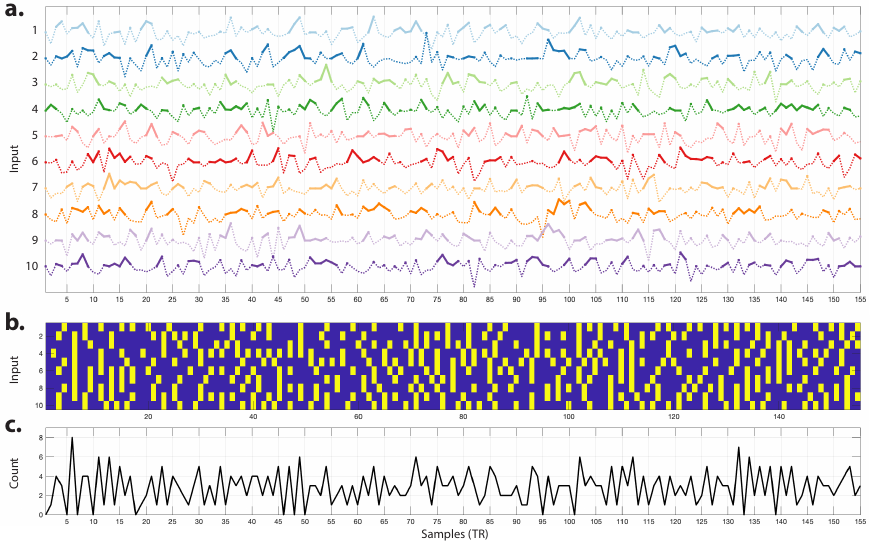}
\caption{\textbf{Sample subject's estimated inputs' temporal profiles.} 
  \textbf{(a)} Color-coded time series shows the temporal profile of the ten inputs during the auditory simulation task in a sample subject during wakeful states. The dashed lines highlight the time points below the inputs' mean. \textbf{(b)} local peaks with above mean values in panel \textbf{a} time series. \textbf{(c)} Total number of identified peaks at each time point in panel \textbf{b}.}
\label{SI_Figure_5}
\end{figure*}

\begin{figure*}
\centering
\includegraphics[width=1\linewidth ]{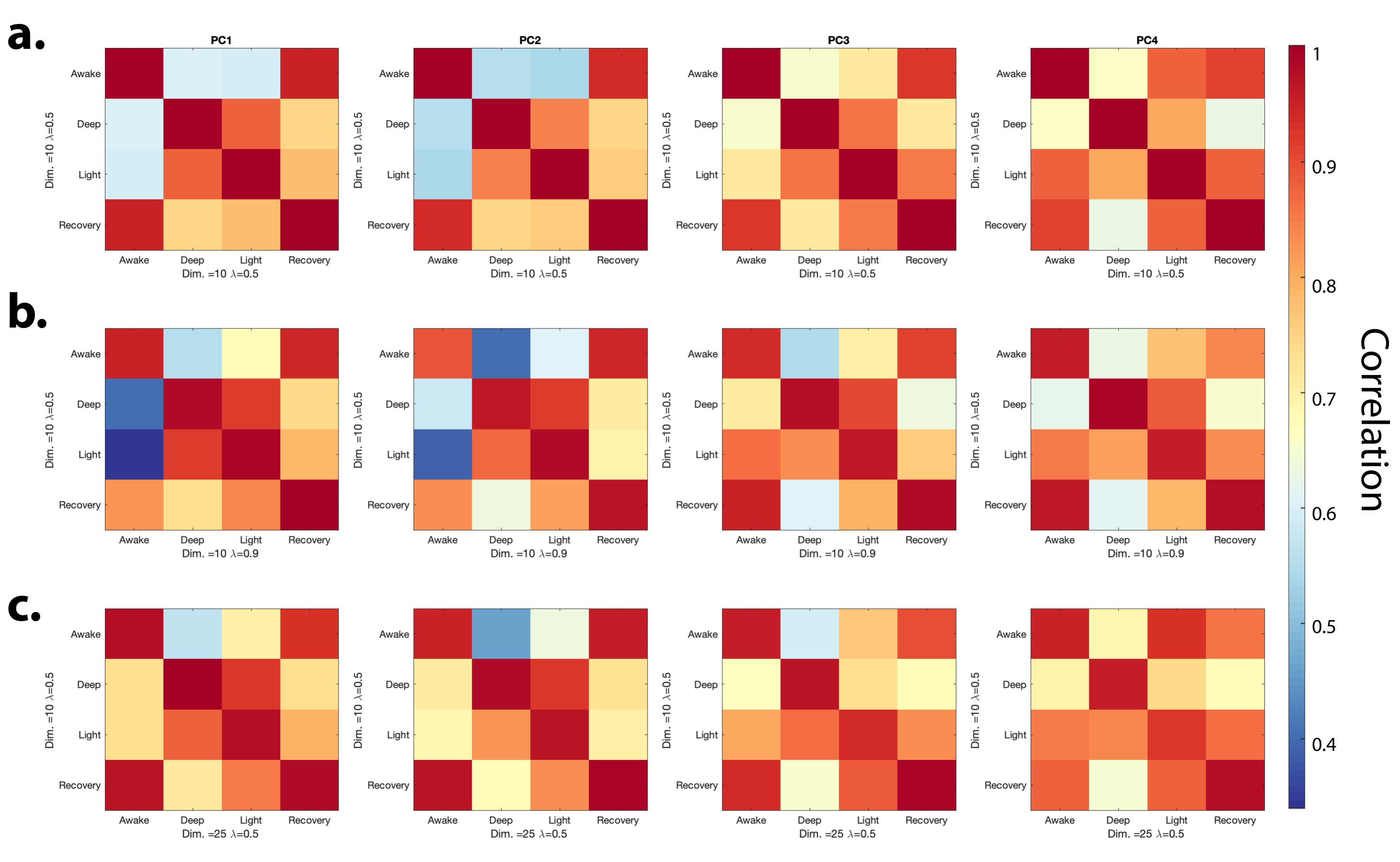}
\caption{\textbf{Comparison of leading input Principal Components (PCs) across hyperparameters} This figure compares the leading principal components (PCs 1-4) identified from the spatial profiles of estimated inputs (matrix B) across different consciousness states. The analysis explores the impact of varying hyperparameters on the extracted PCs.\textbf{(a)} Correlation matrix showing the similarity between PCs identified using an input dimension of 10 and a regularization factor of 0.5. \textbf{(b)} Correlation matrix showing the similarity between PCs identified using an input dimension of 10 and a regularization factor of 0.5 (same as panel \textbf{a}), compared to those identified using an input dimension of 10 and a regularization factor of 0.9. \textbf{(c)} Correlation matrix showing the similarity between PCs identified using an input dimension of 10 and a regularization factor of 0.5 (same as panel \textbf{a}), compared to those identified using an input dimension of 25 and a regularization factor of 0.5.}
\label{SI_Figure_2}
\end{figure*}

\begin{figure*}
\centering
\includegraphics[width=1\linewidth ]{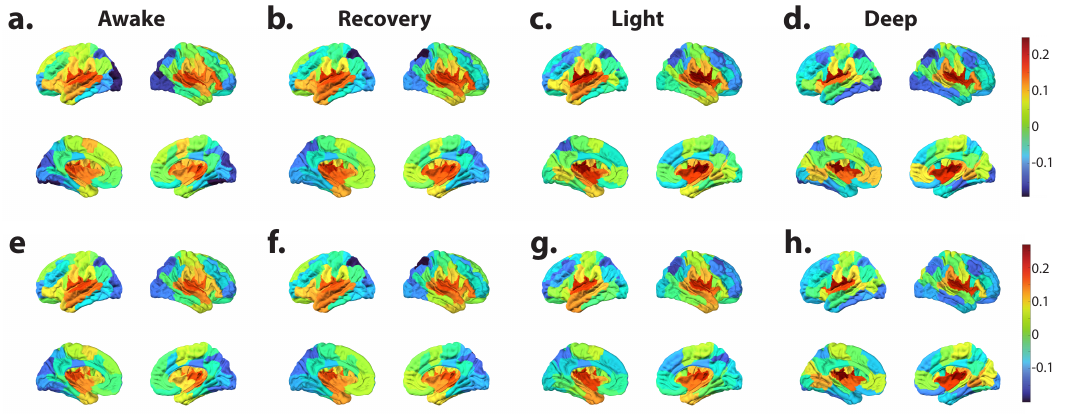}
\caption{\textbf{Principal Component Analysis (PCA) of the Inputs' spatial profiles estimated during auditory stimulation paradigm.} 
\textbf{a-d.} The third principal component (PC3) of the Inputs' spatial profiles, estimated using input dimension = 10 and regularization factor = 0.5 during auditory stimulation. PC3 captures the activation of the auditory cortex and other active regions. \textbf{e-h.} The third principal component (PC3) of the Inputs' spatial profiles, estimated using input dimension = 25 and regularization factor = 0.5 during auditory stimulation. Note that the spread of coactivation to the higher-order auditory cortices in the temporal lobe during the awake and recovery states is captured using both low and high input dimensions}
\label{Auditory_10_25}
\end{figure*}






\end{document}